\documentstyle[preprint,aps,eqsecnum]{revtex}

\newcommand{\be}{\begin{equation}}
\newcommand{\ee}{\end{equation}}
\newcommand{\bea}{\begin{eqnarray}}
\newcommand{\eea}{\end{eqnarray}}
\newcommand{\nn}{\nonumber \\}

\begin{document}
\baselineskip=1.5em
\draft
\preprint{\small CGPG-96/7-1  hep-th/9609009}
\title
{\baselineskip=1.5em General covariance, and supersymmetry \\
without supersymmetry}
\author{Viqar Husain} 
\address
 {\baselineskip=1.5em Center for Gravitational Physics and Geometry,\\ 
Department of Physics, Pennsylvania State University, \\
 University Park, PA 16802-6300, USA\\
 and\\
 Department of Mathematics and Statistics, University of New Brunswick,\\
Fredericton, N. B. E3B 5A3, Canada.}
\maketitle

\begin{abstract}  
\baselineskip=1.5em
An unusual four-dimensional generally covariant and supersymmetric $SU(2)$ 
gauge theory is described. The theory has local degrees of freedom, 
and is invariant under a local (left-handed) chiral supersymmetry, which is 
half the supersymmetry of supergravity.  The Hamiltonian 3+1 decomposition 
of the theory reveals the remarkable feature that the local supersymmetry 
is a consequence of  Yang-Mills symmetry, in a manner reminiscent of how 
general coordinate invariance in Chern-Simons theory is a consequence of 
Yang-Mills symmetry. It is possible to write down an infinite number of 
conserved currents, which strongly suggests that the theory is classically 
integrable. A possible scheme for non-perturbative quantization is outlined.  
This utilizes ideas that have been developed and applied recently to the
problem of quantizing gravity. 
\end{abstract}

\bigskip

\section{Introduction} 

In any theory with local gauge symmetries, the symmetries of the
action are normally manifest in the Hamiltonian theory as first class
constraints on the phase space variables. Standard examples of this
are the Gauss law constraint on the phase space of Yang-Mills theory,
and the space and time reparametrization constraints, called respectively 
the spatial-diffeomorphism and Hamiltonian constraints, on the phase
space of general relativity.

There are however exceptions to this rule. In topological field
theories such as the $SU(N)$ $BF$ theory \cite{bf}, the action 
is invariant under both general coordinate transformations and Yang-Mills 
gauge transformations. However the Hamiltonian theory has {\it more} 
first class constraints than is indicated by these symmetries of the
action. It is these extra constraints that are responsible for
eliminating all local (or propagating) degrees of freedom, leaving only
a finite number of global or topological degrees of freedom. For example, 
for $SU(N)$ $BF$ theory in 3+1 dimensions, there are $4(N^2-1)$ first 
class constraints per space point whereas the configuration space has
$3(N^2-1)$ degrees of freedom per space point. 

Another well known example of this type is Chern-Simons theory, where
for gauge group $SU(N)$, there are $N^2-1$ first class constraints per
point on $2(N^2-1)$ phase space degrees of freedom per point; the usual 
counting after imposing gauge fixing conditions therefore gives 
no local degrees of freedom. 

In general, in $SU(N)$ topological field theories there are {\it more} 
first class constraints on the phase space than the $(N^2-1)+4$ that 
one might at first guess from the $N^2-1$ Yang-Mills and four spacetime 
reparametrization invariances of the action. This happens in the 
$B-F$ and Chern-Simons theories because the field equations imply that 
the Yang-Mills connections is flat.  

There is an example of a four-dimensional theory \cite{vk} in which,
unlike topological theories, there are {\it less} first class
constraints on the phase space than is  evident from an inspection
of the symmetries of the action. The action for this theory, which 
resembles that for general relatvity in the first order Palatini form, 
is 
\be 
S= \int e^{AB}\wedge e_{BC} \wedge F_C^{\ A}, 
\label{action}
\ee 
where $e_\mu^{\ AB}$ is a 1-form and $F(A)=dA+A\wedge A$ is the
curvature of the Yang-Mills connection $A_\mu^{\ AB}$. The Yang Mills
gauge group is $SU(2)$ ($A,B,...=1,2$ are $SU(2)$ 2-spinor indices), and
$\mu\nu,...$ are spacetime indices. This would be an action for
general relativity if the gauge group is taken to be $SL(2,C)$ instead
of $SU(2)$. Then instead of the dreibeins $e_\mu^{\ AB}=e_\mu^{\
(AB)}$, there would be vierbeins $e_\mu^{\ AB'}$ in the action (where
as usual the $A'$ etc. are complex conjugate spinor indices), with no 
other change, that is, the connection would still be $SU(2)$ valued 
\cite{lt,sam}. Such an action leads to the Ashtekar Hamiltonian 
formulation of general relativity \cite{ash}. 

This theory (\ref{action}) is not metric independent; the spacetime metric
is $g_{\mu\nu}=e_\mu^ie_\nu^j\delta_{ij}$. However, beacuse $e_\mu^i$ is 
a dreibein, this metric is degenerate with signature
$(0+++)$. The theory has genuine local degrees of freedom; it is a true field
theory rather than a topological field theory with only a finite
number of physical degrees of freedom. One way to see this is to count
the number of first class phase space constraints: there is the $SU(2)$ 
Gauss law constraint and, as shown in \cite{vk}, {\it only} three other 
constraints which are the spatial diffeomorphism constraints. It turns out 
that the Hamiltonian constraint corresponding to time reparametrizations is 
not an additional constraint, but rather vanishes identically.  The 
configuration space variable is the spatial component of the $SU(2)$ 
connection, so there are $9-3-3=3$ degrees of freedom per spatial point.  
Thus this theory has one more local degree of freedom per point than 
general relativity.

If we wish to quantize this theory, we might first consider 
perturbation theory. However, there is no expansion of the action
(\ref{action}) that isolates a quadratic `free theory' term and a
`non-linear' interaction term, and so one cannot construct a perturbative
quantum theory in the usual way. The same is true for general
relativity in the first order covariant formulation, and it has been
suggested \cite{witten} that this is the essential reason that general
relativity is non-renormalizable as a perturbative quantum
theory. Hence one would conclude, for the same reason, that the  
theory (\ref{action}) is also non-renormalizable.  However it has been shown
that the quantum theory exists non-perturbatively \cite{ab}: There is 
a Hilbert space with well defined operators acting in the space. Thus 
the action (\ref{action}) leads to a completely integrable quantum 
field theory! 

As far as classically integrability is concerned, it is possible to 
write down an infinite number of constants of the motion: Since the 
Hamiltonian constraint vanishes identically, the spatial integral of any 
density constructed from the 3-metric is a constant of the motion; every 
invariant of three-geometry ($\equiv$ three-metrics modulo diffeomorphisms) 
is a  constant of the motion. However no complete Poisson commuting set is 
known that would explicitly prove Liouville integrability.

We turn now to generally covariant field theories with local supersymmetry. 
The only examples of such  theories are supergravity and certain topological 
field theories with super-group Yang-Mills invariance.  In this paper we 
describe an unusual theory of this type which is neither of these, and which 
has a number of interesting properties. 

The theory has local degrees of freedom, and two unusual
features. The first feature is that {\it the local supersymmetry of 
the action is a consequence of its local Yang-Mills symmetry}. This is 
reminiscent of how general covariance in Chern-Simons theory is a 
consequence of Yang-Mills symmetry \cite{witten}. The second is that 
the Hamiltonian constraint corresponding to the time reparametrization 
symmetry is identically zero. This feature is just like that for the 
action (\ref{action}) above. For this reason the theory we discuss 
in this paper may be viewed as a supersymmetric version of (\ref{action}). 

The motivation for studying such a theory, apart from curiosity, is at least 
threefold: (i) Because of the unusual features described above, it is possible 
to write down explicitly an infinite number of constants of the motion, which 
strongly suggests that the theory is integrable. If a proof of 
integrability can be given, say in the sense of presenting the equations 
of motion in a Lax form, it would provide the first 
example of an integrable supersymmetric and generally covariant field 
theory in four dimensions with local degrees of freedom,  (ii) the 
methods used in establishing the existence of the quantum theory of the 
action (\ref{action}) may, with some modification, be used here 
as well to present a complete quantization, and (iii) the fact that the 
spacetime metric is degenerate and the Hamiltonian constraint is 
identically zero means that the theory is in a sense only 
three-dimensional: The initial data given on a three-dimensional 
Cauchy surface does not change. Although the action is a 
four-dimensional one, it is in this sense `already 
dimensionally reduced'. This suggests looking for higher dimensional 
generalizations and string type actions  with the similar features. 

The outline of this paper is as follows: In the next section we give
the action and dicuss its symmetries and field equations. 
Section III contains the  Hamiltonian version of the theory, with an 
explanation of how the supersymmetry is realized at the canonical level.  
Section IV is a discussion of classical observables, and Section V 
contains a brief description of quantization. This is followed by a 
concluding section.

\section{The model}

The theory we consider in this paper is given by the action 
\be 
S = \int_M [\ e^{AB}\wedge e_{BC}\wedge F_A^{\ C} + 
  \alpha \ e^{AB}\wedge \psi_B \wedge D\psi_A\ ], \label{lag}  
\ee
where $e_\mu^{\ AB}$ is a one-form (bosonic) dreibein field, 
$\psi_\mu^A$ is an anticommuting (fermionic) chiral spin 3/2 field,  
and $A_{\mu A}^{\ \ B}$ is the (bosonic) $SU(2)$ gauge field. The 
indices $A,B,...$ are (chiral) two-spinor indices, and the 
covariant derivative and curvature are defined as usual by  
\bea
D \lambda_A = d \lambda_A + A_A^{\ B}\wedge \lambda_B, \nn
F_A^{\ B} = d A_A^{\ B} + A_A^{\ C}\wedge A_C^{\ B}. 
\eea  

The conventions for raising and lowering spinor indices with the
antisymmetric spinor $\epsilon^{AB}$, and its inverse $\epsilon_{AB}$, 
are $\lambda^A = \epsilon^{AB}\lambda_B$ and $\lambda_A = \lambda^B
\epsilon_{BA}$, where the $\epsilon$'s satisfy
$\epsilon^{AC}\epsilon_{BC} = \delta^A_B$. The fields $\psi^A_\mu$, 
being anticommuting, satisfy the conditions 
\be
\psi^A \wedge\psi^B = \psi^B \wedge \psi^A.
\ee
>From this it follows that
\be
\psi^A\wedge \psi_A =0.
\label{spinid}
\ee

This action, being the integral of a four-form, is manifestly
invariant under spacetime diffeomorphisms. It is also invariant 
 under local $SU(2)$ gauge transformations. The action has an 
additional local boson-fermion  symmetry when the coupling constant 
$\alpha=1$. As we now show, this is an on shell local supersymmetry. 
Consider the local transformations
\be  
    \delta_\lambda e_{AB} =  \psi_{(A} \lambda_{B)}\ \ \ \ \ \ \ \ 
    \delta_\lambda \psi_A = - D\lambda_A, 
\label{susy}
\ee
where $\lambda_A(x)$ is an anticommuting (spacetime dependent) 
parameter. These satisfy the (anti-)commutation rules 
\be
[ \delta_\lambda, \delta_\rho ]\ \psi^A = 0,\ \ \ \ 
\{ \delta_\lambda, \delta_\rho   \}\  \psi^A = 0, 
\ee
\bea 
[ \delta_\lambda, \delta_\rho ]\ e^{AB} &=& 
\rho^{(A}\delta_\lambda\psi^{B)} + \lambda^{(A}\delta_\rho\psi^{B)} 
= D(\rho^{(A}\lambda^{B)}),\nn 
\{ \delta_\lambda, \delta_\rho \}\ e^{AB} &=& 
\rho^{(A}\delta_\lambda\psi^{B)} - \lambda^{(A}\delta_\rho\psi^{B)} 
=\rho^{(A}D\lambda^{B)} +  \lambda^{(A}D\rho^{B)}.
\eea
These relations are the left-handed component of 
the supersymmetry transformations in supergravity. 
(See for example \cite{ted}.)    

Under the tranformations (\ref{susy}), the change in the lagrangian is 
\bea 
\delta_\lambda L &=& 2\ (1-\alpha)\ 
\psi^{(A}\lambda^{B)}\wedge e_{BC}\wedge 
F_A^{\ C} +\alpha\ \lambda_A \wedge 
(\ {1\over 2} \psi^B \wedge \psi^A - De^{AB}\ )\wedge  D\psi_B \nn
& &+\  {\rm surface\ terms}.  
\label{delS}
\eea  
For the parameter value  $\alpha =1$ the action is invariant under the 
local supersymmetry transformations (\ref{susy}): The first term in the 
variation above vanishes, and the second term becomes  proportional to 
the equation of motion for $A_A^{\ B}$, which, for $\alpha = 1$, is  
\be
D(e^{BA} \wedge e^C_{\ B}) +  e^{B(A} \wedge \psi^{C)}\wedge \psi_B = 0.  
\label{De}
\ee
By expansion of the symmetrization on the r.h.s., this last equation 
implies  
\be 
De^{AB} = {1\over 2} \psi^A \wedge \psi^B.   
\label{gs}
\ee
Using this, the second term in (\ref{delS}) vanishes. 
This establishes that (\ref{lag}) in invariant under the  local supersymmetry 
generated by (\ref{susy}), modulo the  equation of motion (\ref{De}).  
(It is also the case in supergravity that one of the chiral supersymmetries 
is on shell in exactly this way \cite{ted}. See below.)
 
It is straightforward to make the action invariant under local supersymmetry 
without the use of an equation of motion. This is accomplished by 
extending the supersymmetry transformations (\ref{susy}) to act also on 
the gauge field $A_A^{\ B}$. The necessary transformation on $A_A^{\ B}$ 
may be deduced from the variation of the action. It is non-linear 
(and unconventional), and given by  
\be 
\delta_\lambda A_{AC} \wedge e^C_{\ B}=
-{1\over 2} \lambda_{(A} D\psi_{B)}.   
\ee 

The  lagrangian for our model is similar to the following first order 
(complex) lagrangian for supergravity \cite{ted}, which is made from only the 
self-dual part of the spin connection: 
\be
S_{SUGRA} =  \int_M [\ i\ e^{AA'}\wedge e_{A'B}\wedge F_A^{\ B} - 
 e^{AA'}\wedge \bar{\psi}_{A'} \wedge D\psi_A\ ].
\label{sugra}
\ee
Here the fields $e_\mu^{\ AA'}$ are vierbiens rather than dreibeins, 
and both the unprimed $\psi_\mu^A$ and their complex conjugate primed 
$\psi_\mu^{A'}$ spinor fields are present. This supergravity action is 
separately invariant under left- and right-handed supersymmetry 
transformations, with  invariance under the right-handed part 
being modulo the equations of motion for $A_\mu^{\ AB}$. These  
transformations are, respectively,
\be
\delta_\lambda e_{AA'} = -i \bar{\psi}_{A'} \lambda_{B},\ \ \ \ \ \ \ \ 
    \delta_\lambda \psi_A = 2D\lambda_A, \ \ \ \ \ \ \ \ 
    \delta_\lambda \bar{\psi}_{A'} = 0, 
\label{left}
\ee 
and 
\be
\delta_{\bar{\lambda}} e_{AA'} = -i\psi_A \bar{\lambda}_{A'},\ \ \ \ \ \ \ \ 
    \delta_{\bar{\lambda}}\bar{\psi}_{A'} = 2D\bar{\lambda}_{A'}, \ \ \ \ \ \ \ 
\ 
    \delta_{\bar{\lambda}} \psi_{A} = 0,
\label{right}
\ee 
together with $\delta_\lambda A^{AB} = \delta_{\bar{\lambda}} A^{AB}=0$. 
There is therefore an additional `right-handed' supersymmetry 
transformation in supergravity which is absent in our  model. It is this 
additional transformation which, when anti-commuted with the left-handed 
transformation, gives the usual spacetime translation generator in 
supergravity. Also, the number of bosonic and fermionic fields in 
supergravity, $e^{AA'}$ and the pair ($\psi^A$,${\bar{\psi}}^{A'}$), are 
equal in number, whereas in our model there is a mismatch with  twelve 
$e^{AB}$ and only eight $\psi^A$. While this is unusual, the 
transformations (\ref{susy}) are manifestly still a local 
boson-fermion symmetry. For comparison, we note that the 
supersymmetry in the heterotic string is similar - the supersymmetry 
generator is a chiral spinor \cite{gsw}. We note also that this is not 
the so called `$\kappa$-supersymmetry' \cite{gsw}, which has the property 
that there is no associated conserved  Noether charge. The variation of the 
action under the chiral supersymmetry above does lead to a non-trivial
Noether charge. This is also reflected in the  Hamiltonian theory below,  
in the fact that there is a first class constraint associated with the 
 supersymmetry. 
 
A further difference from supergravity is that there is a special 
spacetime direction in our theory which arises essentially 
because there is a dreibein field in four dimensions. This direction 
is given by the vector density 
\be
\tilde{u}^\alpha = \epsilon^{\alpha\beta\gamma\delta}e_\beta^{\ AB} 
e_{\gamma B}^{\ \ C} e_{\delta AC}, 
\ee
where $ \epsilon^{\alpha\beta\gamma\delta}$ is the metric independent
Levi-Civita tensor density, and is orthogonal to $e_\mu^{\ AB}$:
$\tilde{u}^\mu e_{\mu AB} =0$. There is a corresponding special
  2-spinor density  given by  
\be 
\tilde{\phi}^A =  \tilde{u}^\mu \psi^A_\mu.
\ee 
 
The spacetime metric is $g_{\mu\nu} = e_\mu^{\ AB}e_{\nu BA}$ 
and is degenerate with signature $(0+++)$, where the degeneracy 
direction is $\tilde{u}^\mu$. This situation is identical to the 
non-supersymmetric theory given in \cite{vk}.     

The other equations of motion following from (\ref{lag}), obtained by 
varying $e_\mu^{AB}$ and $\psi_a^A$  (with $\alpha=1$), are 
respectively 
\bea
2\ e_{C(B}\wedge F_{A)}^{\  C} + \psi_{(B} \wedge D\psi_{A)} = 0, \\
\psi_B \wedge De^{AB} =0. \label{delpsi}
\eea
We notice that this last equation (\ref{delpsi}) is identically satisfied 
when  the equation of motion for $A_a^{\ AB}$ (\ref{gs}) holds, together 
with the spinor identity (\ref{spinid}). As we will see in the next section, 
it is due to this fact that the supersymmetry turns out to be a consequence 
of Yang-Mills symmmetry. 
 
\section{Hamiltonian theory} 

We now consider the Hamiltonian formulation of the theory considered
in the last section. Assuming that the 4-manifold $M = \Sigma\times R$,
where $\Sigma$ is a spatial 3-manifold and $R$ is `time', it is
straightforward to rewrite the action (\ref{lag}) in a $3+1$ form.
This gives
\bea 
S = & & \int_R dt \int_\Sigma d^3x\ \bigl[\ E^{cA}_{\ \ B}
\dot{A}_{aA}^{\ \ B} 
+ \Pi^{aB} \dot{\psi}_{aB} 
+ e_0^{\ AB}\epsilon^{abc}(\ 2e_{aBC}F_{bcA}^{\ \ \ C} 
- \psi_{aB}D_b \psi_{cA}\ ) \nn 
&+& A_{0A}^{\ \ C}(\ D_aE^{aA}_{\ \ C} + \Pi^{aA}\psi_{aC}\ ) 
+\psi_{0A}(\ \epsilon^{abc}e_a^{\ AB}D_b\psi_{cB} - D_a\Pi^{aA}\ )
\bigr],
\label{3+1a}
\eea 
where $a,b,...$ are there dimensional tensor indices, $e_0^{\ AB},
A_{0A}^{\ \ B}$ and $\psi_{0A}$ are the time components of the various
fields, and 
\be 
\epsilon^{abc}: = \epsilon^{0abc} 
\ee
 is the 3-dimensional Levi-Civita tensor density. The variables 
\be 
E^{aA}_{\ \ B} := 2\epsilon^{abc}e_b^{\ AC}e_{cCB}\ \ \ \ {\rm and} 
\ \ \ \ \Pi^{aA} := -\epsilon^{abc}e_b^{\ AB}\psi_{cB},
\label{mom}
\ee
are the canonical momenta conjugate to the configuration variables
$A_{aA}^{\ \ B}$ and $\psi_{aA}$ respectively. Thus the phase space is
parametrized by the boson variables 
$(A_{aA}^{\ \ B},\ E^{aA}_{\ \ B})$ and the fermionic variables 
$(\psi_{aA},\ \Pi^{aA})$.  It is instructive to write down the 
fundamental equal time Poisson brackets for the fermionic variables: 
\be
\{ \epsilon^{abc}e_b^{\ AB}\psi_{cB}(x), \psi_{dD}(y) \} 
= \delta^a_d \delta^A_D \delta^3(x,y). 
\ee
Since $e_a^{AB}$ is effectively the variable canonically conjugate to 
$A_a^{\ AB}$, three combinations of the spinor and space components of 
$\psi_{aA}$ are canonically conjugate to three other of its components. 
Therefore the fermionic part of the configuration space is effectively 
coordinatized by three functions and not six. \footnote{I would like to 
thank Ted Jacobson for clarifying this point.} (An analagous feature 
appears in Hamiltonian Chern-Simons theory, where the components of the 
Yang-Mills connection are canonically conjugate to one another.) 
To indicate that only half the functions in $\psi_a^A$ are configuration 
coordinates, we set   
\be 
\psi_a^A = \phi_a \delta^A_1, \ \ \ \ \ \Pi^{aA} = \pi^a \delta^A_1  
\ee
for all that follows, where $\phi_a$ and $\pi^a$ are anticommuting 
variables. 

Since the triads $e_a^{AB}$ are assumed to be 
non-degnerate, the density 
\be 
e:={1\over3!}\epsilon^{abc}e^A_{aB}\ e^B_{bC}\ e^C_{cA}\ne 0,
\ee
and the canonical variables $E^{aA}_{\ \ B}$ are dual to the 
triads $e^A_{aB}$:
\be 
E^{aA}_{\ \ B}=e e^{aA}_{\ \ B},
\ee
with
\be 
e^{aA}_{\ \ B}e^B_{aC}=\delta^A_C\ \ {\rm and}\ \ 
e^{aA}_{\ \ B}e^B_{bA} = \delta^a_b.
\ee 

The variation of the $3+1$ action (\ref{3+1a}) with respect to the 
non-dynamical time components of the fields $e_0^{\ AB}, A_{0A}^{\ \ B}$ 
and $\psi_{0A}$, leads to the three constraints 
\bea
G^A_{\ B}:= D_aE^{aA}_{\ \ B} + \Pi^{aA}\psi_{aB} = 0,\label{gauss} \\   
\epsilon^{abc}[\ 2e_{aC}^{\ \ (B}F_{bc}^{A)C} 
- \psi_a^{(B}D_b \psi_c^{A)}\ ] = 0, \\
\epsilon^{abc}e_a^{\ AB}D_b\psi_{cB} - D_a\Pi^{aA}=0,
\label{sucon1}
\eea
where the symmetrization on the spinor indices in the second equation is 
due to $e_a^{\ AB} = e_a^{\ BA}$.

As expected in a generally covariant theory, the Hamiltonian is a
linear combination of constraints, with $e_0^{\ AB}, A_{0A}^{\ \ B}$
and $\psi_{0A}$ as the lagrange multipliers.
On general grounds we expect that there will be constraints on the
phase space associated with every local symmetry of the action.
Therefore there should be four constraints associated with the 3-space
and time reparametrization invariance, two constraints associated with
the `right-handed' local supersymmetry transformations (\ref{susy}),
and finally three more constraints associated with the $SU(2)$
Yang-Mills invariance. To see if this expection is borne out, we must
first rewrite the constraints as functions of only the canonical
variables, and exhibit their Poisson algebra. 

The first constraint (\ref{gauss}) is the usual Gauss law associated
with the $SU(2)$ Yang-Mills invariance, and it is already a function of
only the phase space variables. 

The second constraint may be rewritten as a function of the 
phase space variables by multiplying it by $\epsilon^{dab}e_d^{\ AB}$ 
and then using third constraint. The result is 
\be 
E^{aA}_{\ \ C}F_{abA}^{\ \ \ C} - \Pi^{aA}D_b\psi_{aA} + 
\partial_a(\Pi^{aA}\psi_{bA}) =0. 
\ee
By adding to this a term proportional to the Gauss law constraint, 
specifically $A_{aA}^{\ \ B}\times G^A_{\ B}$, it becomes the 
spatial diffeomorphism constraint
\be 
C_b := E^{aA}_{\ \ C}\partial_a A_{bA}^{\ \ C} 
- \partial_a( E^{aA}_{\ \ C}A_{bA}^{\ \ C} ) 
- \Pi^{aA}\partial_b\psi_{aA} + \partial_a(\Pi^{aA}\psi_{bA}) = 0. 
\ee 

Finally, using the equation 
\be 
\epsilon^{abc}e_a^{AB}= { E^{[bA}_{\ \ C}E^{c]CB}\over \sqrt{E} }, 
\ee
where $E=\epsilon_{abc}E^{aA}_{\ \ B} E^{bB}_{\ \ C} E^{cC}_{\ \ A}$, 
the third constraint may also be rewritten as a function 
of the phase space variables:  
\be 
S^A:={ E^{[bA}_{\ \ C}E^{c]CB}\over \sqrt{E} }
D_b\psi_{cB} - D_a\Pi^{aA}=0.
\ee
As expected, this constraint generates supersymmetry transformations 
on the phase space variables: With 
\be 
S(\lambda):=\int_\Sigma \lambda_AS^A, 
\ee
we have, for example, 
\be 
\delta_\lambda \psi_A := \{ \psi_A, S(\lambda) \} = -D\lambda_A,   
\ee
which is one of the supersymmetry transformations (\ref{susy}) above.
 
We now point out a rather surprising feature of this supersymmetry 
constraint, namely, that {\it the supersymmetry constraint is 
identically satisfied as a consequence of the $SU(2)$ Gauss law}. To 
show this, we first note from (\ref{sucon1}) that  
\be 
S^A:=\epsilon^{abc}D_b(e_a^{\ AB}\psi_{cB}) 
- \epsilon^{abc}\psi_{cB}D_b e_a^{\ AB} 
 - D_a\Pi^{aA} = - \epsilon^{abc}\psi_{cB}D_b e_a^{\ AB} = 0,
\label{nsusy}
\ee
where we have used the definition of the momentum conjugate to
$\psi_{aA}$ (\ref{mom}). This last form of the supersymmetry 
constraint (\ref{nsusy}), which is just the spatial projection of 
the field equation (\ref{delpsi}), and the Gauss law (\ref{gauss}) 
may be succinctly written as\footnote{We are now writing these 
constraints as 3-forms rather than as scalar densities. The 
difference is multiplication by the Levi-Civita density 
$\epsilon^{abc}$.}
\be
S^A = \psi_B \wedge g^{AB}=0, 
\label{sucon}
\ee 
\be
G^{AB} = 2 e_B^{\ (A}\wedge g^{B)C}=0,
\label{gs2}
\ee
where the 2-form $g^{AB}$ is defined by 
\be 
g^{AB} := De^{AB}-{1\over 2} \psi^A\wedge \psi^B.
\ee 
Now, it can be shown directly by expansion from (\ref{gs2}) that  
\be 
 g^{AB}_{[ab]} = e^{cA}_{\ \ C}G^{CB}_{[abc]} 
- {\epsilon^{def}\over \sqrt{E}} e^{AB}_{[a}G^{CD}_{b]ef} e_{dCD}.
\label{g(G)}
\ee
Therefore we have 
\be 
G^{AB}=0 \Longleftrightarrow  g^{AB}=0, 
\ee
a result which was asserted above in equation (\ref{gs}).

It is now clear that the third constraint, may be written  
as a function of the Gauss law by using (\ref{g(G)}) in  
(\ref{sucon}). This gives the explicit formula   
\be
S^A = \psi_B \wedge (\ e^{cA}_{\ \ C}G^{CB}_{[abc]} 
- {\epsilon^{def}\over \sqrt{E}} e^{AB}_{[a}G^{CD}_{b]ef} e_{dCD}\ )
\ dx^a\wedge dx^b.
\ee
 
Thus we have shown that only the first two of the constraints that 
follow from the action are independent. The third
constraint, which as we saw above generates supersymmetry
transfomations, turns out to be identically satisfied when the $SU(2)$
Gauss law holds. This is just the reflection in the 
Hamiltonian theory of the fact that the equation of motion 
(\ref{delpsi}) is identically satisfied as a consequence of (\ref{gs}) . 
 Thus there are effectively only the spatial
diffeomorphism and Gauss law constraints in the Hamiltonian theory.

Putting all the above observations together, the final Hamiltonian form 
of the action, written entirely in terms of the phase space variables and
appropriate lagrange multiplier functions, becomes
\be
S= \int_R dt\int_\Sigma d^3x\ \bigl[\ E^{cA}_{\ \ B}\dot{A}_{aA}^{\ \ B}
-N^aC_a - \mu_A^{\ C}G^A_{\ C} - \lambda_A^{\ C}G^A_{\ C} \bigr],
\label{ha}
\ee
where   
\bea
N^a &:=& e^{aA}_{\ \ B}e_{0A}^{\ \ B},\\
\mu_A^{\ C}&:=& A_{0A}^{\ \ C} - N^a A_{aA}^{\ \ C},\\
\lambda_A^{\ C}&:=& \psi_{0B}\ \bigl( \psi_{aA}e^{aBC} - 
{\epsilon^{abc} \over \sqrt{E}}\ e_a^{DB}e_{bA}^{\ \ C}\psi_{cD}
\bigr). 
\eea

The Hamiltonian equations of motion are obtained by varying this action 
with respect to the canonical variables $E^{cA}_{\ B}$, ${A}_{aA}^{\ \ B}$, 
and the   variables $N^a$, $\mu_A^{\ C}$, and 
$\Lambda_A^{\ C}$. Varying these gives the two constraints 
\be 
C_a=0=G_A^{\ B}.
\ee 
Varying the canonical variables gives the Hamiltonian equations of 
motion
\be 
\dot{E}^{aA}_{\ \ B} = \{ E^{aA}_{\ \ B}, H \}, \ \ \ 
\dot{A}_{aA}^{\ \ B} = \{ A_{aA}^{\ \ B}, H  \},
\ee
where the Hamiltonian is 
\be 
H = \int_\Sigma d^3x\ \bigl[\ N^aC_a + \mu_A^{\ B}G^A_{\ B} + 
\lambda_A^{\ B}G^A_{\ B}\ \bigr]. 
\ee

Since $C_a$ is the generator of spatial diffeomorphisms, and 
$G^A_{\ B}$ the generator of Gauss rotations, the evolution 
equations are simply 
\be 
\dot{A}_{aA}^{\ \ B} = {\cal L}_N A_{aA}^{\ \ B} 
 + D_a\mu_A^{\ B} + D_a\lambda_A^{\ B},
\ee
and 
\be
\dot{E}^{aA}_{\ B} =  {\cal L}_N \dot{E}^{aA}_{\ B} 
 + \mu^A_{\ C} {E}^{aC}_{\ B}  + \lambda^A_{\ C} E^{aC}_{\ B} 
\ee
where ${\cal L}$ denotes the Lie derivative.  We therefore see that
`evolution' in this theory amounts to spatial diffeomorphisms, and a
pair of Gauss rotations of the canonical variables. In particular,  
{\it the supersymmetry of the covariant action manifests itself in 
the Hamiltonian theory only as a second Gauss transformation.}

Our model does for local supersymmetry what Chern-Simons theory
does for general coordinate invariance: Local supersymmetry 
is a consequence of local Yang-Mills symmetry 
just as general coordinate invariance in Chern-Simons theory is a 
consequence of local Yang-Mills symmetry. In both these cases, the 
generators of the respective symmetries are functions of the Gauss 
law generator. 
  
While we have shown this directly at the Hamiltonian level, it 
may be shown at the covariant level as well by appropriately 
constructing the Gauss transformation function out of the physical 
fields and an arbitrary grassmann variable. A fundamental difference 
from Chern-Simons theory is, of course, that our model has local 
 degrees of freedom - there are six independent 
configuration degrees of freedom per point. 

Apart from the supersymmetry being a consequence of Yang-Mills symmetry, 
the theory has another unusual feature: The Hamiltonian constraint, 
which is the generator of time reparametrization invariance, does not 
appear in the Hamiltonian action (\ref{ha}). This feature is already 
present in the absence of grassmann fields, and has been discussed 
in detail in Ref. \cite{vk}. 

A comparison of the Hamiltonian theory of this model with Hamiltonian 
supergravity, (for example the one derived from the action (\ref{sugra})), 
provides a further understanding of the supersymmetry and spacetime 
reparametrization constraints. On the phase space of supergravity there are 
two first class constraints asssociated with the left and right-handed 
supersymmetry transformations (\ref{right}-\ref{left}). The Poisson 
bracket of these two constraints yields, in one guise or another 
\cite{d'eath,fradkin,teit}, the generator of spacetime reparametrizations - 
the Hamiltonian and spatial-diffeomorphism constraints. In spinorial 
variables, these last two constraints arise together as one constraint 
with a pair of left and right-handed spinor indices, in the form 
${\cal H}_{AA'} =0$. It is in this way that the supersymmetry generators 
close to give the spacetime reparametrization generators. By contrast, 
in our model there is only the left-handed supersymmetry and the time 
reparametrization constraint vanishes identically. The supersymmetry 
constraint turns out to be proportional to the Gauss constraint, 
indirectly giving closure of the supersymmetry constraint algebra. 
(This is one way of seeing that the supersymmetry transformations close.) 

\section{Observables}  

There are different points of view about what is an observable in 
a generally covariant theory. An observable in any theory 
with first class constraints is defined to be a phase space function(al) 
 which Poisson commutes with all the first class constraints. This 
basically gives  the gauge invariant phase space variables, whose 
dynamics may then be studied using the Hamiltonian of the theory. 
In a generally covariant theory, the dynamics is  generated 
by a constraint because the theory is invariant under time 
reparametrizations. If we apply the above  prescription for 
finding the  observables, we would be seeking  constants of  
motion, which in fact do not evolve. This has led to suggestions that 
observables in a generally covariant theory should Poisson commute 
with only the `kinematical' first class constraints. 

In our model,  the Hamiltonian constraint 
vanishes identically so there is no issue about how to define observables:  
These are phase space functionals that Poisson commute with the Gauss and 
spatial-diffeomorphism constraints. Such functionals are also constants of  
motion. It is easy to write down an infinite number of them - the spatial 
integral of any Gauss law invariant scalar density of weight one. These 
are naturally divided into three classes: Those involving only the 
gravitational variables, only the fermionic variables, or both. The `electric' 
3-metric made from the dreibein $E^{aAB}$, and the `magnetic' 3-metric 
made from $H^{aAB} = \epsilon^{abc} F_{bc}^{\ AB}$ may be used to construct 
the   scalar curvatures $R(E)$ and $R(H)$. (We are assuming invertibility 
of the respective 3-metrics.) Examples 
of each of the three types of observables are 
\bea
&& \int_\Sigma d^3x\  ({\rm det} E) R(H), \\
&& \int_\Sigma d^3x\  \Pi^{Aa}\psi_{aA}, 
\eea
and 
\be
 \int_\Sigma d^3x\  \Phi(R(H),R(E))\ \Pi^{Aa}\psi_{aA},
\ee
where $\Phi$ is an arbitrary function of the curvature scalars of the 
electric and magnetic metrics. There are clearly an infinite number of such 
examples. 

There are also diffeomorphism invariant loop observables which are constructed 
by defining loops using the phase space variables, rather than introducing loops 
as auxilliary variables \cite{vloops}. This is done by defining loops as the 
intersections of two 2-dimensional surfaces, where the surfaces themselves 
are defined as the level surfaces of scalar fields made from the phase space 
variables. One example is the loop $\gamma(c_1,c_2)$ defined by setting 
\be f(R(E),R(H))=c_1 \ \ \ \ \ \ g(R(E),R(H))=c_2,  \ee
where $f,g$ are arbitrary functions of the Ricci scalars of the electric and 
magnetic metrics. This loop is constructed using  only the gravitational 
variables. Then, the Wilson loop 
\be  W[\gamma,E,A] = {\rm Tr P exp}\int_{\gamma(c_1,c_2)} dx^a A_a ,
\label{wloop}
\ee  
based on such a `matter loop', is a constant of the motion. 
One can do similar things using the spinor variables $\Pi^{aA}$ and $\psi_a^A$, 
and also define loop observables with insertions along the loop \cite{vloops}. 
This  gives  spatial-diffeomorphism invariant versions of the Rovelli-Smolin 
loop variables for canonical gravity \cite{rs}. 

The standard criteria for integrability is an algorithm for generating  
an infinite number of Poisson commuting constants of the motion. 
For 2-dimensional theories, or theories which can be effectively written 
as 2-dimensional theories such as self-dual gravity, such an algorithm is 
provided by the Lax, or zero-curvature form, of the evolution equations. 
In the absence of a similar procedure here, the fact that we can write down 
an infinite number of constants of motion is only suggestive that this model 
is integrable. 

\section{Quantization} 

In this section we briefly consider the quantization of the model described 
in the preceding sections. As we have already pointed out, the action 
does not lead to a clear separation of  `free theory'  and 
`perturbation' terms. Therefore there does not seem to be a way to construct 
a perturbative quantum field theory starting from the action (\ref{lag}). 
We therefore consider non-perturbative Dirac quantization. A possible choice 
of representation for the quantum theory is the connection representation, 
where the wavefunctionals are $\Phi[A_{aA}^{\ \ B}, \psi_a^A]$. There are 
two sets of  Dirac quantization conditions. 
\bea
G^A_{\ B} |\Phi> &=&   [\ D_a\ {\delta\over \delta A_{aA}^{\ \ B}}  
+ \psi_a^A \ {\delta\over \delta \psi_a^B}\   ]\  
\Phi[A_{aA}^{\ \ B}, \psi_a^A] = 0, \\
C_b |\Phi> &=&   [\ ( \partial_a A_{bA}^{\ \ B} )\ 
{\delta\over \delta A_{aA}^{\ \ B}}
- \partial_a(A_{bA}^{\ \ B}\ {\delta\over \delta A_{aA}^{\ \ B}} ) \nn
&\ & -  ( \partial_b\psi_a^A  )\ {\delta\over \delta \psi_a^A} 
+ \partial_a( \psi_b^A\ {\delta\over \delta \psi_a^A} ) \ ] \ 
\Phi[A_{aA}^{\ \ B}, \psi_a^A]= 0.
\eea
The first condition states that $|\Phi>$ is invariant under $SU(2)$ gauge 
transformations {\it and} under the left-handed supersymmetry transformations. 
This is because,  as explained in the preceding sections, the latter are a 
consequence of the former. The second states that $|\Phi>$ is invariant under 
spatial diffeomorphisms. This is a remarkably simple prescription for obtaining 
the quantum states. Furthermore, since  all the quantum constraints are linear 
in the momenta, there is no operator ordering ambiguity, and the quantum 
constraint 
algebra closes in  the same way as the classical Poisson algebra. 

It is a straightforward exercise to write down any number 
of quantum states: These are those observables of the last section which are 
functionals of only the $A_{aA}^{\  \ B}$ and $\psi_a^A$. A class of purely 
bosonic states are traces of the Wilson loops, for loops $\gamma(c_1,c_2)$  
defined by $f(R(H))=c_1,\ g(R(H))=c_2$. There are no purely fermionic states 
because it is not possible to form a non-zero scalar density using only 
$\psi_a^A$. Mixed states may be constructed by using the spinor density 
\be 
\tilde{\chi}_A= \epsilon^{abc} H_{aAB}H_b^{\ BC}\psi_{cC},  
\ee
to form the scalar density $\chi = 
\sqrt{\epsilon_{AB}\tilde{\chi}^A\tilde{\chi}^B}$. 
A class of mixed states is then 
\be 
\phi = \int_\Sigma d^3x\ \chi\ f(R(H),\sqrt{\chi}),
\ee 
where $f$ is an arbitrary function of its arguments. 
It is straightforward to produce many other examples. What is lacking is a 
systematic way of constructing a Hilbert space, which in turn is connected 
with producing a closed infinite dimensional algebra of physical observables. 

A systematic approach for constructing a quantum theory may be to find a 
suitable generalization of the methods of Ashtekar et. al. in \cite{ab}, which 
were developed for application to diffeomorphism invariant theories of 
connections, 
such as general relativity in the Ashtekar formulation \cite{ash}. These methods 
have already been used to show that the theory given in \cite{vk}, (which is the 
bosonic action (\ref{action}) above), is an integrable quantum field theory. 
To see the form that such a generalization might take, we first give a brief 
outline of the main steps used in this approach, which is applicable to 
theories where the only configuration space variable is a connection: 
\begin{list}
 {}
 \item 1. The Gauss law invariant states are functions $\Psi(\bar{A})$ on the 
space of generalized connections modulo gauge transformations, where the 
generalization is a suitable enlargement of the classical configuration space - 
the space of {\it smooth} connections modulo gauge transformations.  
Generalized connections may be distributional as well as smooth. 

\item 2. There is an innerproduct, and an orthonormal `spin network' basis on 
the space of generalized connections which is labelled by closed graphs 
 \cite{ab,baez,clspin}. Associated with each 
edge of the graph is a matrix which is the holonomy of $\bar{A}$ in a fixed 
representation (`color'). Associated with each vertex of the graph is an 
`intertwiner' matrix,  which ties up all the matrix indices on the edges 
meeting at that vertex. This gives gauge invariance. Any finite number of edges 
can meet at a vertex. (These spin networks are a generalization of Penrose's 
spin networks \cite{pen}). 

\item 3. The diffeomorphism constraint is implemented on this space, using 
this basis, via 
`exponentiation'. A unitary operator representing finite diffeomorphisms can 
be defined. Using this operator, diffemorphism invariant states are constructed 
by `integrating over the group'. This gives an infinite class of quantum states. 
There is a natural inner product on the space of diffeomorphism invariant 
states obtained in this way. 
\end{list}

To apply a similar procedure to the present model requires incorporating  
the Rarita-Schwinger spinor fields $\psi_a^A$, which are now a part of the 
configuration space. 

The first step, which produces Gauss law invariant spin network states, may be 
 generalized by allowing  insertions of the $\psi_a^A$ at the vertices, 
along with the intertwiner matrices. A class of open graphs  also give gauge 
invariant states. These are the graphs whose open ends are the ends of edges in 
the fundamental representation of $SU(2)$. Such ends can be plugged with a 
$\psi_a^A$ to give gauge invariance. The simplest such graph is one edge in 
the fundamental representation, with a $\psi_a^A$ at each end. 

It appears, at least at first sight, that the second step  goes through as well. 
The orthonormality of the spin network states comes from `integration over 
the connection', which is really a group integration over the holonomies 
associated 
with the edges of graphs. Therefore insertions of  $\psi_a^A$'s on the vertices 
do not effect these integrations. What they do effect are the degeneracies -   
the number of graphs with a fixed number of edges and representations, but with 
the 
indices at the vertices tied up in different ways. A major difference, of 
course, 
is that the spin network states now also carry space indices (since the 
$\psi_a^A$'s do), which may make implementation of the third step more 
difficult. 

\section{Discussion}

We have described a model in which  supersymmetry appears in a slightly 
different light - as a consequence of Yang-Mills symmetry. For this reason, 
perhaps  it should not be called supersymmetry at all, at least at the 
Hamiltonian 
level.  However, at the level of the action  there is manifestly a chiral 
supersymmetry. A concomitant feature of this constraint structure is that the 
Hamiltonian constraint vanishes identically, which is to be contrasted with 
supergravity, where the left- and right-handed supersymmetry constraints close 
on the spacetime reparametrization constraints. It is the fact that the time 
reparametrization constraint vanishes identically that allows us to write down 
an infinite number of constants of motion, and also gives a possible 
interpretation 
of the theory as one that is already dimensionally reduced. From the point of 
view 
of the Hamiltonian evolution equations, the only change in the initial data 
under 
evolution is that due to spatial-diffeomorphisms and Yang-Mills gauge 
transformations. 

It would be worthwhile to see if the  quantization scheme suggested above can be 
carried to completion. If so, it would provide the first example of an 
integrable 
four-dimensional supersymmetric quantum field theory, as well as a concrete way 
to 
introduce matter into the methods developed for pure connection theories. 

\acknowledgements

I would like to thank Abhay Ashtekar, Murat G$\ddot{\rm u}$naydin, Ted Jacobson 
and 
Lee Smolin for helpful remarks, and Jack Gegenberg and Jon Thompson for 
hospitality 
at the University of New Brunswick, where this work was completed. This work was 
supported by NSF grant PHY 93-96246, the Eberly Research Funds of the 
Pennsylvania 
State University, and by the Natural Science and Engineering Research Council 
of Canada.

\end{document}